\documentclass[twocolumn, prl, aps, prl, floats,floatfix,amsmath,amssymb, longbibliography, secnumarabic, superscriptaddress,preprintnumbers]{revtex4-1} 
\usepackage[final]{graphicx}
\usepackage{amsmath}
\usepackage{bbm}
\usepackage{bm}
\usepackage{amsfonts}
\usepackage{amssymb}
\usepackage{latexsym}
\usepackage{graphicx}
\usepackage[english]{babel}
\usepackage{multirow}
\usepackage{float}
\usepackage{url}
\usepackage{slashed}
\usepackage{xcolor} 
\usepackage[utf8]{inputenc}
\usepackage{stmaryrd} 
\usepackage{enumitem}
\usepackage{hyperref}
\usepackage{cleveref}
\usepackage{siunitx}
\usepackage{verbatim}

\newcommand{\be}{\begin{equation}}
\newcommand{\ee}{\end{equation}}
\newcommand{\ba}{\begin{array}}
\newcommand{\ea}{\end{array}}
\newcommand{\bea}{\begin{eqnarray}}
\newcommand{\eea}{\end{eqnarray}}

\newcommand{\besub}{\begin{subequations}}
\newcommand{\eesub}{\end{subequations}}
\newcommand{\gev}{{\ \rm GeV}}
\renewcommand{\eqref}[1]{Eq.~(\ref{eq:#1})}

\definecolor{darkerblue}{rgb}{0.2,0.2,0.5}
\definecolor{seagreen}{rgb}{0.180392,0.545098,0.341176}

\begin{document}

\title{Re-examining the Solar Axion Explanation for the XENON1T Excess}

\author{Christina Gao}
\email{yanggao@fnal.gov}
\affiliation{Theoretical Physics Department, Fermi National Accelerator Laboratory, Batavia, IL, 60510, USA}

\author{Jia Liu}
\email{liuj1@uchicago.edu}
\affiliation{Enrico Fermi Institute, University of Chicago, Chicago, IL 60637, USA}

\author{Lian-Tao Wang}
\email{liantaow@uchicago.edu}
\affiliation{Enrico Fermi Institute, University of Chicago, Chicago, IL 60637, USA}
\affiliation{Department of Physics, University of Chicago, Chicago, IL 60637, USA}

\author{Xiao-Ping Wang}
\email{xia.wang@anl.gov}
\affiliation{HEP Division, Argonne National Laboratory, 9700 Cass Ave., Argonne, IL 60439, USA}

\author{Wei Xue}
\email{weixue@ufl.edu}
\affiliation{Department of Physics, University of Florida, Gainesville, FL 32611, USA}

\author{Yi-Ming Zhong}
\email{ymzhong@kicp.uchicago.edu}
\affiliation{Kavli Institute for Cosmological Physics, University of Chicago, Chicago, IL 60637, USA}

\preprint{EFI-20-13}

\begin{abstract}
The XENON1T collaboration has observed an excess in electronic recoil events below $5~\mathrm{keV}$ over the known background, which could originate from beyond-the-Standard-Model physics. The solar axion 
is a well-motivated model that has been proposed to explain the excess, though it has tension with astrophysical observations.  
The axions traveled from the Sun can be absorbed by the electrons in the xenon atoms via the axion-electron 
coupling. Meanwhile, they can also scatter with the atoms through the inverse
Primakoff process via the axion-photon coupling, which emits a photon and mimics the electronic recoil signals. We found that the latter process cannot be neglected.  After including the $\rm{keV}$ photon produced via inverse Primakoff in the detection, the tension with the astrophysical constraints can be significantly reduced. We also explore scenarios  involving additional new physics to further  alleviate the tension with the astrophysical bounds. 
\end{abstract}
\maketitle

Axions are pseudo-goldstone bosons which  naturally arise from the beyond-the-Standard-Model (BSM)
physics scenarios \cite{Peccei:1977hh, Weinberg:1977ma, Wilczek:1977pj}. Due to an approximate shift symmetry, they can be naturally light. 
Typically, they are very weakly coupled to other particles, 
which makes them a good candidate of dark matter  or dark sector particles \cite{Preskill:1982cy, Abbott:1982af, Dine:1982ah}. The phenomenology of the axions is rich and they give unique signals in cosmology, 
astrophysics, and particle physics \cite{Raffelt:1990yz, Duffy:2009ig, Kawasaki:2013ae, Marsh:2015xka, Graham:2015ouw}.

XENON1T, a dual-phase Liquid Xenon detector, is one of the leading experiments looking for dark matter. 
Due to its large volume and low backgrounds, the XENON1T is also sensitive to other rare processes potentially 
related to the BSM physics. Recently, the XENON1T collaboration reported their searches for the
low-energy electronic recoil, with an excess in the range of 1-5$\,\rm{keV}$, which cannot be accounted for  by the 
known backgrounds~\cite{Aprile:2020tmw}. The XENON1T collaboration has also  performed a fit to the excess using the solar axion model \cite{vanBibber:1988ge}. Since this report, 
there have been active speculations about the explanation of the excess \cite{Takahashi:2020bpq,OHare:2020wum,Kannike:2020agf,Amaral:2020tga,Alonso-Alvarez:2020cdv, Fornal:2020npv,Boehm:2020ltd,Harigaya:2020ckz,Bally:2020yid,Su:2020zny,Du:2020ybt,DiLuzio:2020jjp,Bell:2020bes,Chen:2020gcl,AristizabalSierra:2020edu,Buch:2020mrg,Choi:2020udy,Paz:2020pbc,Dey:2020sai,Khan:2020vaf, Cao:2020bwd,Primulando:2020rdk, Nakayama:2020ikz, VanTilburg:2020jvl,  Lee:2020wmh,Graciela:1802691,Yongsoo:1802,1802727,1802729, Smirnov:2020zwf}.

It is tempting to explain the  XENON1T excess using the solar axions since the axion energy spectrum naturally matches the excess.  
The axions are produced in the Sun from several processes, including
the Primakoff process $\gamma \, + \, Ze \, \to \,  Ze \, + \, a$;
the Atomic axion-recombination and de-excitation, Bremsstrahlung, and Compton scattering processes~(ABC); and the nuclear transitions. Hence, the axion-photon $g_{a\gamma}$, axion-electron $g_{ae}$ and axion-nucleon
$g_{an}$ couplings enter the production.
With its tiny coupling to photons, the $\rm{keV}$ axions have a long lifetime  and can travel from the Sun to the XENON1T.
For the processes in the detector which can give the signal, XENON1T~\cite{Aprile:2020tmw} considered only the axion-electron coupling. In this case,  the axions could be absorbed
by the electrons in xenon atoms. 

The relevant axion couplings can be summarized in the following Lagrangian,   
\begin{align}
\mathcal{L} \supset - g_{ae}  \frac{\partial_\mu a}{2 m_e}  \bar{e} \gamma^\mu \gamma_5 e - \frac{1}{4} g_{a\gamma} a F_{\mu\nu} \tilde{F}^{\mu\nu} .
\label{eq:axion}
\end{align}
${F}^{\mu\nu}$ is the field strength of photon, and its dual $\tilde{F}^{\mu\nu} = 
\frac{1}{2}\epsilon^{\mu\nu \alpha \beta} F_{\alpha \beta}$.
However, the parameter space of the solar axion interpretation of the excess is in tension with he 
astrophysical observations of stellar evolution including the White Dwarfs (WD) and the Horizontal Branch (HB) stars in 
the globular clusters (GC)~\cite{Aprile:2020tmw,DiLuzio:2020jjp}. 

\begin{figure}[htb]
   \includegraphics[width=0.9\columnwidth]{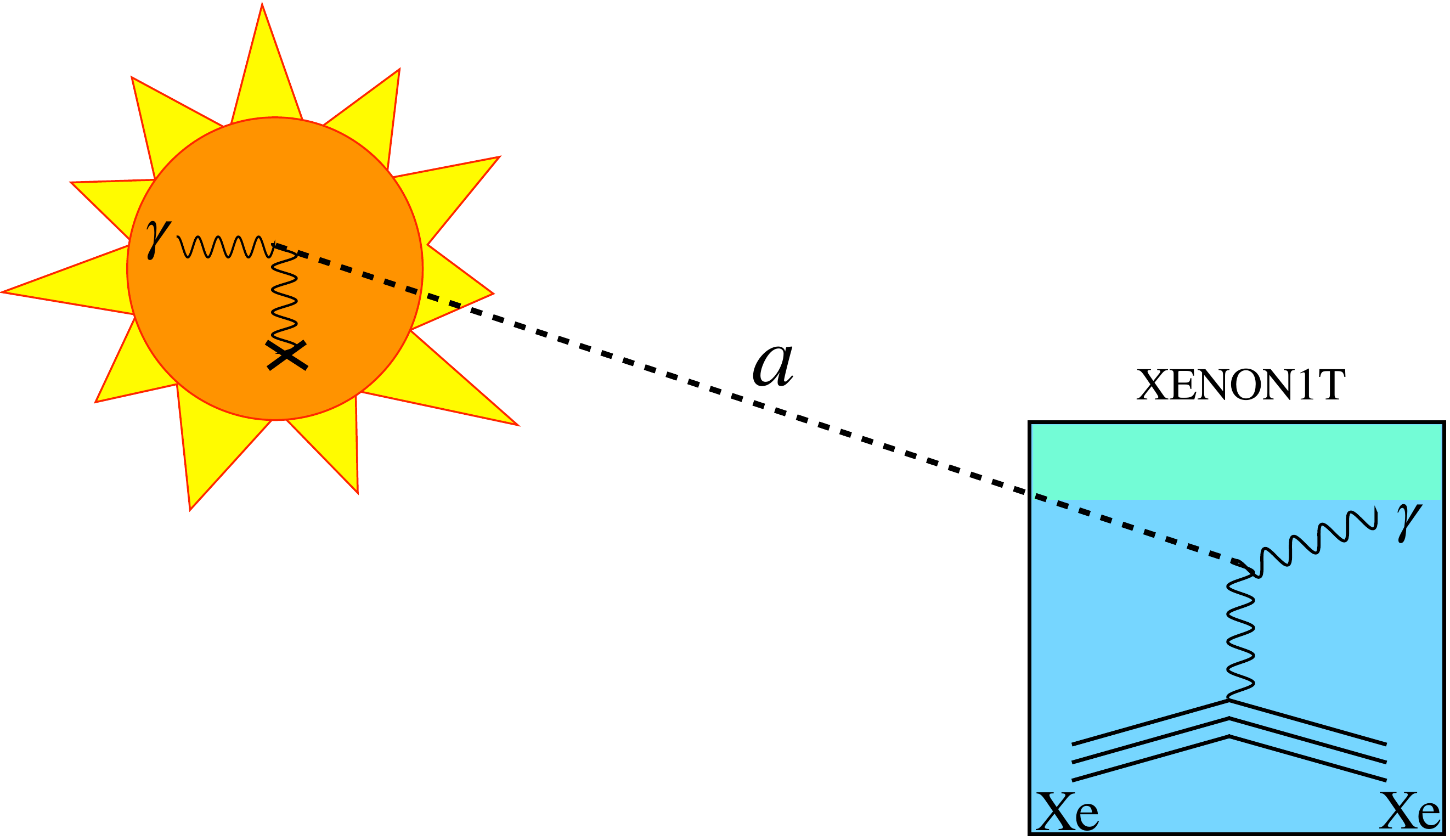}
   \caption{The solar axion induced photon signal through the inverse Primakoff process. 
   }
   \label{fig:process}
\end{figure}

In this letter, we take into account the fact that at $\mathrm{keV}$ energy range, the current XENON1T experiment can hardly distinguish the detector response of photons from that of electronic recoils. Hence, instead of electronic recoil, 
the low-energy photons generated through the inverse Primakoff scattering between the solar axion and the xenon atoms in the detector can mimic the electronic signal, as shown in Fig.~\ref{fig:process}. 
Using inverse Primakoff process to detect axion was proposed in the cryogenic experiments via Bragg scattering \cite{Buchmuller:1989rb, Paschos:1993yf, Creswick:1997pg}, and was applied by the SOLAX, COSME, CUORE, CDMS and EDELWEISS collaborations \cite{Avignone:1997th, Morales:2001we, Arnaboldi:2002du, Arnaboldi:2003tu, Ahmed:2009ht, Armengaud:2013rta}. However, it was not included in the liquid  time projection chamber type of experiments previously.   
We show that, after including both the electronic recoil and the inverse Primakoff process, the tension between the solar axion explanation and the astrophysical constraints is significantly reduced.

The letter is structured as follows: we first describe the detection using the inverse Primakoff process, and after considering the astrophysics and terrestrial constraints, we present the fit to the data of XENON1T. We then discuss the possible extensions of new physics to further alleviate the tension between the constraints and the XENON1T fit. We conclude in the end.
\\

\noindent \textit{\textbf{Detection from inverse Primakoff process.}}---
In this section, we compute the contribution to the electronic recoil  from the inverse Primakoff process
$
a + \rm Xe \to \gamma + \rm Xe,
$
where Xe represents the xenon nucleus. 
 The differential cross section is given by~\cite{Buchmuller:1989rb, Raffelt:1996wa, Creswick:1997pg}:
\begin{align}
\frac{d\sigma^{\rm invPrim}_{a\to\gamma}}{d\Omega} = \frac{\alpha}{16 \pi} g_{a\gamma}^2 \frac{\bm{q}^2}{\bm{k}^2}\left(4 -\bm{q}^2/\bm{k}^2 \right) F_a^2(\bm{q}^2),
\label{eq:diffsigma}
\end{align}
where $\alpha$ is the fine structure constant, $\bm{k}$ is the momentum of the incoming axion and $\bm{q}$ is the momentum transfer. In the limit of small axion mass, $m_a \ll |\bm{k}|$, the energy of the outgoing photon is also approximately $|\bm{k}|$. The form factor $F_a$,  characterizing the screening effect of the electric charge of the nucleus, is taken to have the following ansatz:
$
 F_a (\bm{q}^2)= {\mathcal N} Z \bm{k}^2/(r_0^{-2} + \bm{q}^2).
$
The atomic number of xenon $Z = 54$. $r_0$ is the screening length~\cite{Buchmuller:1989rb} and $\mathcal N$ represents a normalization factor. They can be determined numerically by fitting the form factor above to the expression $F_a = \bm{k}^2/\bm{q}^2(Z- F_\gamma)$~\cite{Buchmuller:1989rb}, where the atomic form factor $F_\gamma$ is reported in Ref.~\cite{ITC}. From the two-parameter fit, we obtain $\mathcal N = 0.54$, and $r_0^{-1} = 6.31 {\rm \ keV} = (31.3 {\rm \ pm})^{-1}$ which is close to the reciprocal of the xenon atomic radii $108$ pm \cite{doi:10.1063/1.1712084}. This screening length corresponds to a screened charge of $Z_{sc} = 5.3$ for xenon at $|\bm q| = 3$ keV.

Next, we calculate the event rate from solar axions with both the inverse Primakoff process and the axioelectric effect. The cross section of the latter process is given by \cite{Pospelov:2008jk, Alessandria:2012mt}
\begin{align}
\sigma_{\rm ae} = \sigma_{\rm pe} \frac{g_{\rm ae}^2}{\beta_a} \frac{3 E_a^2}{16\pi \alpha m_e^2} 
\left(1-\frac{\beta_a^{2/3}}{3}\right),
\end{align}
where $\sigma_{\rm pe}$ is the photoelectric cross-section \cite{Arisaka:2012pb} and $\beta_a$ is the axion velocity.
We will focus on the low energy excess ($\lesssim 5$ keV) throughout this letter, hence only consider the contributions to solar axion flux from
the ABC process,  $\Phi_a^{\rm ABC}$,  and the Primakoff process,  $\Phi_a^{\rm Prim} $, and neglect that from nuclear transition of $^{57}$Fe.
The ABC flux originates from the axion-electron coupling and is given by
$\Phi_a^{\rm ABC}\propto g_{ae}^2$ \cite{Redondo:2013wwa}.
The Primakoff flux is given by \cite{book1}
\begin{align}
	& \frac{d\Phi_a^{\rm Prim}}{d E_a} = 6\times 10^{10} \text{cm}^{-2} \text{s}^{-1} \text{keV}^{-1} \times \nonumber \\ &\left(\frac{g_{a\gamma}}{10^{-10}\text{GeV}} \right)^2 \left(\frac{E_a}{\text{keV}} \right)^{2.481}
	e^{-E_a/(1.205 \text{keV})} .
\end{align}

Given the solar axion flux $\Phi_a$, 
the differential event rate after including both axioelectric and inverse Primakoff processes in the detection is given by
\begin{align}
\frac{dR}{d E_r}= \frac{N_A}{A}&\left(\frac{d \Phi^{\rm ABC}_a}{dE}(E_r) +\frac{d\Phi^{\rm Prim}_a}{dE}(E_r) \right)\nonumber\\
&\times\left(\sigma^{\rm invPrim}_{a\to\gamma}(E_r)+\sigma_{ae}(E_r)\right) ,
\end{align}
where $N_A$ is Avogadro constant, $A =131$ is the atomic weight of xenon, and $E_r$ represents the electronic recoil energy, which is faked by photons in the inverse Primakoff process. 

To compare with the results reported by the XENON1T collaboration, we further smear the differential event rate with a Gaussian with its variance satisfying $\sigma/E_r=(a/{\sqrt{E_r}}+b)\%$.
A numerical fit to the data of XENON1T energy resolution~\cite{XENON:2019dti} yields $a=35.9929$ keV$^{1/2}$ and $b=-0.2084$. After the smearing, we apply the detector efficiency~\cite{Aprile:2020tmw}. 

\begin{figure}[htb]
    \includegraphics[width=0.9\columnwidth]{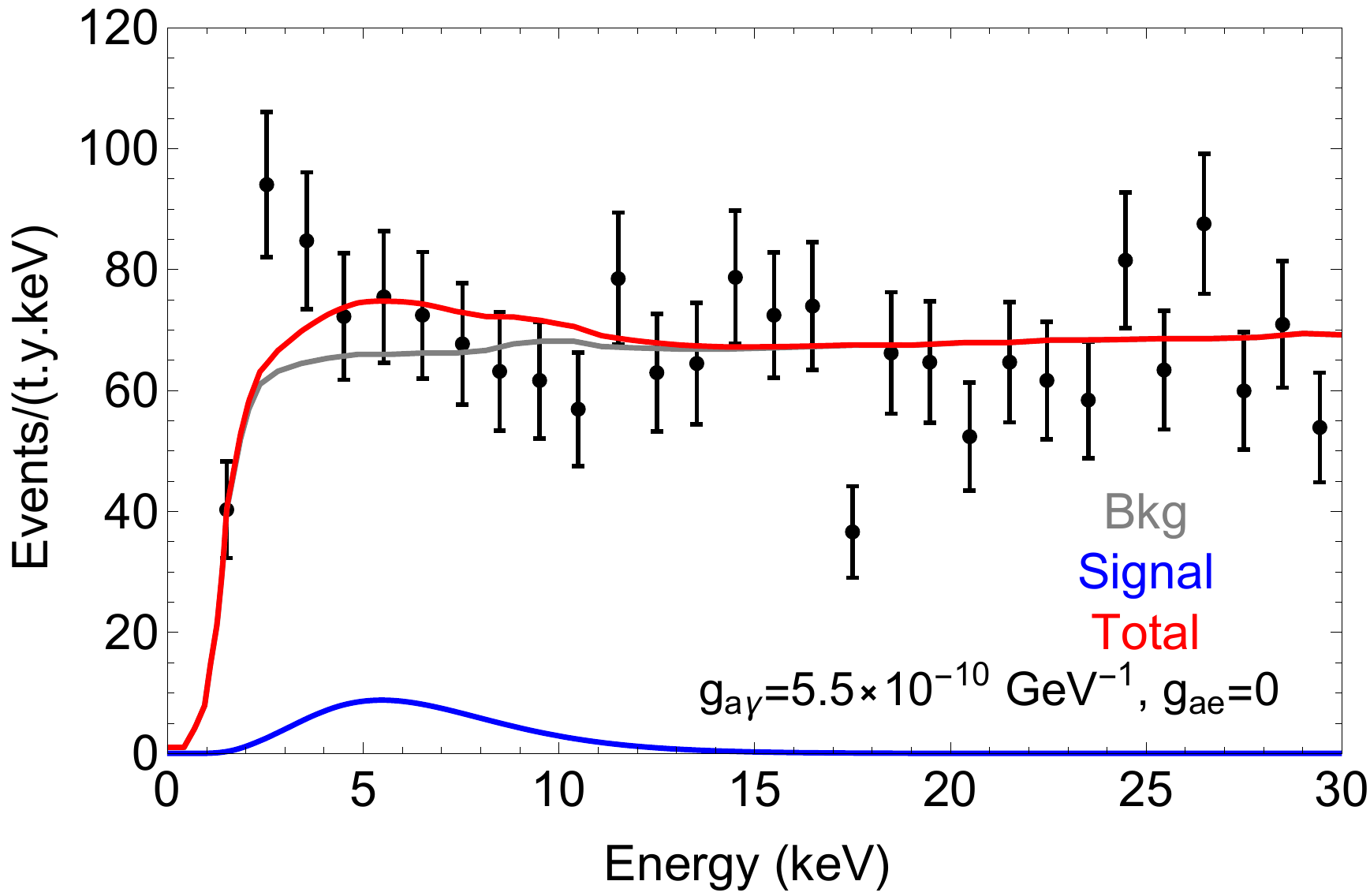}
    \includegraphics[width=0.9\columnwidth]{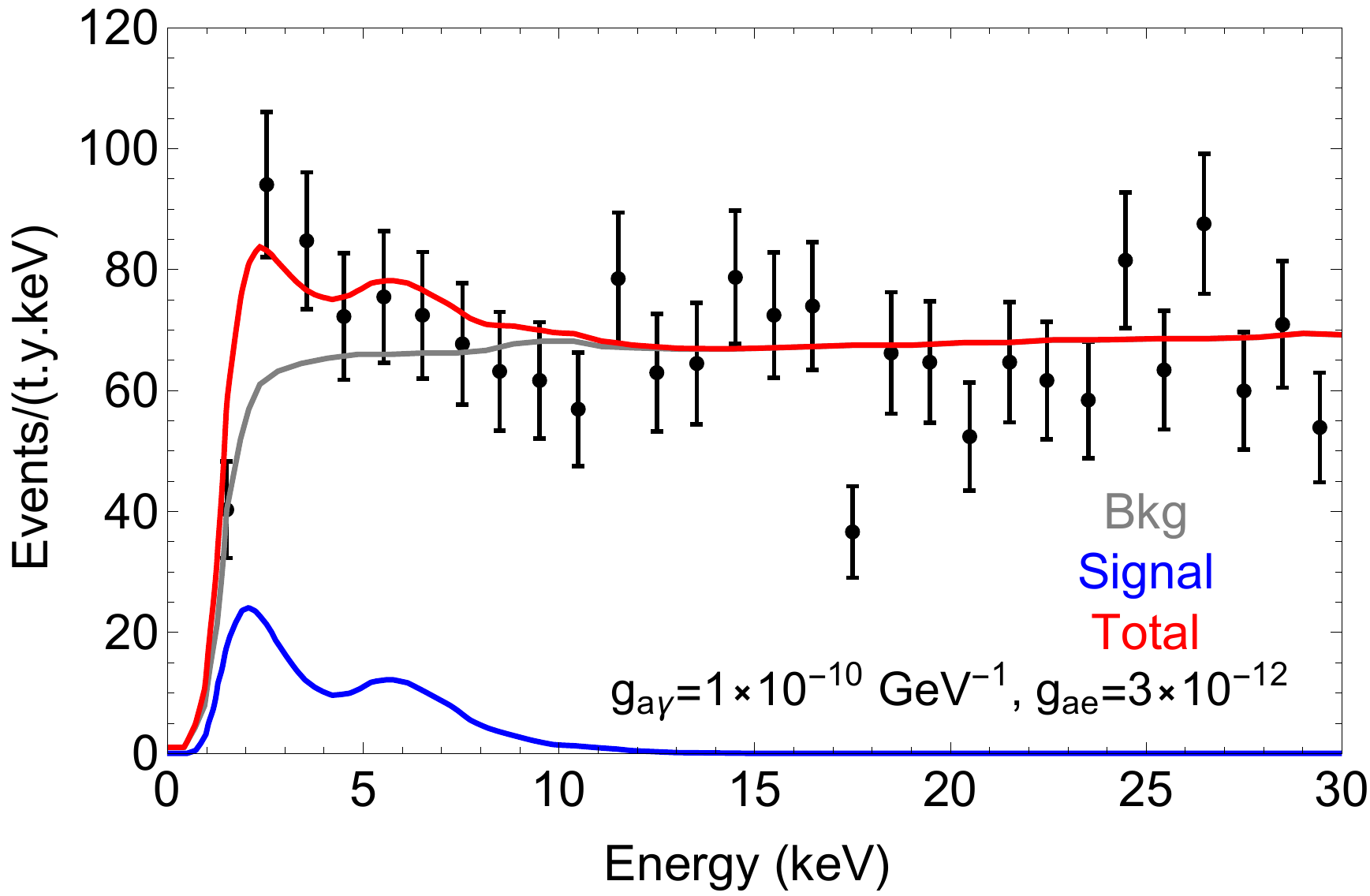} 
   \caption{Fit to electronic recoil energy spectrum with $g_{a\gamma}$ only (top) and both $g_{a\gamma}$ and $g_{ae}$ allowed (bottom).}
   \label{fig:benchmark}
\end{figure}

Fig.~\ref{fig:benchmark} shows two examples of the differential event rate of the electronic recoils given different values of $g_{ae}$ and $g_{a\gamma}$. 
In the case that $g_{ae}=0$, the spectrum is only determined by the detection of $\Phi_a^{\rm Prim}$ through the inverse Primakoff process. It is clear that with $g_{ae}$ switched off, solar axions can still account for the low energy excess, although the fit is not as good as that allowing both $g_{ae}$ and $g_{a\gamma}$ to be non-zero.
\\

\noindent \textit{\textbf{Constraints from astrophysics and terrestrial experiments.}}---
The most severe constraints on the solar axion explanation of the XENON1T excess are from the  stellar cooling in the HB and red-giant branch (RGB) stars, which we review below.

\begin{figure}[h!]
   \includegraphics[width=0.9\columnwidth]{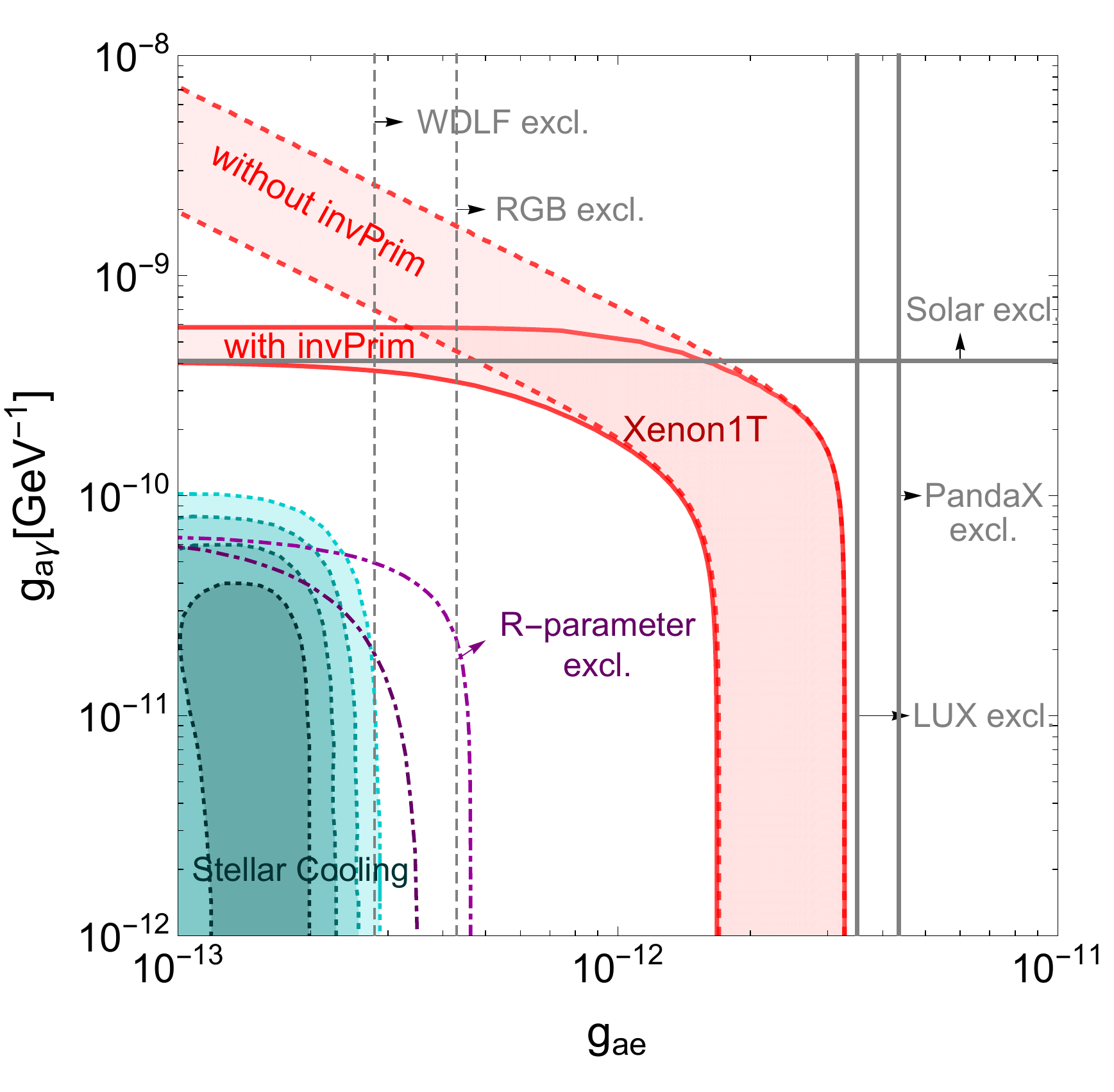}
   \caption{The 2D axion couplings parameter fit for the XENON1T excess after including the inverse Primakoff process. Our best fit (90\% C.L.) to the XENON1T excess is shown in the red shaded region with the solid boundary. In comparison, a ``XENON-like" analysis with only the electron recoil included as the signal yields is shown in the region with the dashed boundary. The main difference is that the inclusion of the inverse Primakoff process allows for a region in which $g_{a \gamma} $ is relatively large while $g_{ae}$ can be small, reducing the tension with the astrophysical data.
   Also included are the constraints (95\% C.L.) from astrophysical observables including WDLF~\cite{Bertolami:2014wua}, the tip of RGB~\cite{Viaux:2013lha} and the $R$-parameter (with two models)~\cite{Giannotti:2015kwo}, as well as the constraints from the global fit of the solar data~\cite{Vinyoles:2015aba}, LUX~\cite{Akerib:2017uem}, and PandaX \cite{Fu:2017lfc}, with arrows denoting excluded regions. We do not show constrains from CAST~\cite{Anastassopoulos:2017ftl} and magnetic WD~\cite{Dessert:2019sgw} by assuming the axion mass $\gtrsim 1\,\text{eV}$. 
 The shaded green region contains 1 $\sigma$ to 4 $\sigma$ contours favored by the anomalous stellar cooling~\cite{Giannotti:2017hny, DiLuzio:2020jjp}. 
   }
   \label{fig:fit}
\end{figure}

Axions with sizable $g_{a\gamma}$ and $g_{ae}$ couplings speed up the burning of the He-core  (H-core) for HB (RGB). The lifetime of the stars in the two phases is proportional to their observed numbers. Therefore, one can use the $R$-parameter, the ratio of the number of HB stars to that of RGB stars, to constrain
the axion couplings. Ref.~\cite{Ayala:2014pea} 
obtained a weighted average $R_\text{av} = 1.39 \pm 0.03$ from the $R$-parameters of 39 low-metallicity galactic GC reported by \cite{Salaris:2004xd} . 
Assuming $g_{ae}=0$, $g_{a\gamma}$ is constrained to be $g_{a\gamma} < 6.6\times 10^{-11}~\text{GeV}^{-1}$. For non-zero $g_{ae}$, Ref.~\cite{Giannotti:2015kwo} presented two stellar evolution models which give slightly different predictions of the $R$-parameter.
In Fig.~\ref{fig:fit}, we adopted the resulting 95\% C.L. constraints on $g_{ae}-g_{a\gamma}$ plane for both models from Fig.~4 of~\cite{Giannotti:2015kwo}.  In the Supplemental material, we further discuss the bound dependence on the He abundance of GCs. The bremsstrahlung energy loss from the axion-electron coupling affects the white dwarf luminosity function (WDLF) and constrains $g_{ae} \lesssim2.8 \times 10^{-13}$~\cite{Bertolami:2014wua}. The same argument on RGB constrains $g_{ae} \lesssim 4.3 \times 10^{-13}$~\cite{Viaux:2013lha}. The global fit of the solar data constrains $g_{a\gamma} < 4.1\times 10^{-10} ~\text{GeV}^{-1}$~\cite{Vinyoles:2015aba}. Other constraints such as X-ray observations on magnetic WDs~\cite{Dessert:2019sgw} get significantly weakened for axion mass  $\gtrsim 1\,\text{meV}$. In Fig.~\ref{fig:fit}, we also show $1\sigma$ to $4 \sigma$ $g_{ae}-g_{a\gamma}$ contours favored by the anomalous stellar cooling~\cite{Giannotti:2017hny, DiLuzio:2020jjp}.

On the terrestrial experiments side, the axion searches from LUX~\cite{Akerib:2017uem} suggest $g_{\rm ae} < 3.5\times 10^{-12}$. Similar constraint is also shown by PandaX \cite{Fu:2017lfc}.
The CAST experiment~\cite{Anastassopoulos:2017ftl} constrains light axions with $g_{a\gamma}< 6.6\times 10^{-11} \gev^{-1}$. This  bound gets significantly weakened for axions with mass $\gtrsim1~\text{eV}$. 
\\

\noindent \textit{\textbf{Results.}}---
In Fig.~\ref{fig:fit}, we present our fit to the XENON1T excess and compare it with the bounds from the previous section. We scan two parameters $g_{ae}$, $g_{a\gamma}$, and apply the method of least squares to the XENON1T data to find the 90\% C.L. contours with (solid red) and without (dashed red) including the inverse Primakoff process. In comparison, we also show the constraints (95\% C.L.) from astrophysical observables including WDLF, the tip of RGB, and the $R$-parameter (with two models), as well as the constraints from the global fit of the solar data and the direct search at LUX and PandaX. The constraints from CAST and magnetic WD can be evaded by the axion with a mass $\gtrsim 1\,\text{eV}$ and we do not show them in Fig.~\ref{fig:fit}.

From Fig.~\ref{fig:fit}, we see that the inclusion of the inverse-Primakoff process has a significant impact on the parameter region preferred by the XENON1T data. In particular, it opens up a parameter region in which $g_{a\gamma} \gg g_{ae}/\text{GeV}$ and the inverse Primakoff process gives rise to the observed signal. Moreover, for $g_{ae}\sim 10^{-13}$, it prefers a $g_{a\gamma}$ which is a few $\times 10^{-10}\,\text{GeV}^{-1}$, one order of magnitude smaller than the preferred $g_{a\gamma}$ without the inclusion of the inverse Primakoff process, satisfying the constraints from the global fit of the solar data, 
and significantly reducing the tension with the stellar cooling bounds. Future terrestrial axion experiments, such as the International Axion Observatory~\cite{Armengaud:2014gea} and the Multilayer Optical Haloscope~\cite{Baryakhtar:2018doz}, can complement the astrophysical probes and cover the relevant $g_{a\gamma}$ coupling region for axions with mass of several eV.
\\

\noindent \textit{\textbf{Possible extensions.}}---
Even though the inclusion of the inverse Primakoff process can significantly improve the prospect of explaining the XENON1T excess with the solar axions, it is still in tension with the stellar cooling bounds, 
with a discrepancy as large as 8$\sigma$, as claimed by~\cite{DiLuzio:2020jjp}. If the excess is indeed completely due to new physics, there remains three possibilities. It could certainly come from other new physics instead of the solar axion, in which case a new explanation of the keV scale needs to be found. It is also possible that there is additional uncertainty in the stellar cooling bound which still has not been appreciated (see e.g.~\cite{Raffelt_1991, Gratton:2010zk} and the discussion in the Supplemental Material~\footnote{See Supplemental Material for the bound dependence on 
the He abundance of GCs, which includes Refs.~\cite{Ayala:2014pea, Giannotti:2015kwo, Tanabashi:2018oca, Serenelli_2010}}). Instead of pursuing these avenues, we will explore a third possibility, with  new physics in addition to the solar axion. In particular, we focus on the parameter space given by $g_{a\gamma} \gg g_{ae}/\text{GeV}$, where the most relevant constraint is from the $R$-parameter, the cooling from HB. We introduce a $U(1)_B$ gauge boson and discuss its effects.

Consider an 
axion coupling to both photon and dark gauge boson $A'$ 
carrying the $U(1)_B$ Baryon charge,
\begin{align}
\mathcal{L} \supset  - \frac{1}{2} g_{a\gamma A'} a F'_{\mu\nu} \tilde{F}^{\mu\nu} + g_{B} A'_{\mu} J_{\rm B}^\mu \ .
\label{eq:u1b}
\end{align}
The $U(1)_B$ $A'$ couples to the baryonic current $J_B$, but not directly to electrons.  Hence, processes mediated by $A'$ does not suffer the screening effect. 
Therefore,
the Primakoff production in the Sun is increased if $A'$ is lighter than the thermal photon in the plasma, 
and the detection cross-section is enhanced by
 $A^2/Z_{sc}^2\sim 600$ for light $A'$ with $m_{A'}<r_0^{-1}\sim 4\,\text{keV}$.

 The Primakoff cross section is given by
\begin{align}
   \sigma_{\gamma\to a}^{A' \text{Prim}} = \frac{g_{a\gamma A'}^2 \alpha_B A^2 }{4} \left(\frac{2\eta^2 +1}{4\eta^2}\ln\left(4\eta^2+1\right)-1\right)
   \, F_N^2 \ ,
   \label{eq:u1bprod}
\end{align}
where $\eta = |\bm{k}|/m_{A'}$, $\bm{k}$ is the momentum of the axion, $\alpha_B \equiv {g_B^2}/{4\pi}$, 
and $F_N$ is the nuclear form factor 
which is approximately 1 for a momentum transfer of a few keV. 
We follow Ref.~\cite{Raffelt:1996wa} to calculate the Primakoff energy loss due to Eq.~\ref{eq:u1b} for $A'$ with $m_{A'} = 0.1~ (1) \,  \rm{keV}$.  The resulting solar energy loss rate per unit volume is 
\begin{equation}
Q_a^{A' \rm Prim} (\text{Sun}) \approx Q_a^{\rm Prim} (\text{Sun}) \times 16.9 (4.3) \times \frac{\alpha_B}{\alpha} \frac{g_{a\gamma A'}^2}{g_{a\gamma}^2},
\end{equation}
and that for HB  is 
\begin{equation}
Q_a^{A' \rm Prim} (\text{HB}) = Q_a^{\rm Prim}  (\text{HB})\times 15.6 (8.0) \times \frac{\alpha_B}{\alpha} \frac{g_{a\gamma A'}^2}{g_{a\gamma}^2},
\end{equation}
where $Q_a^{\rm Prim}$ is the energy loss rate per unit volume from~Eq.~\ref{eq:axion}.
The cross-section for inverse Primakoff detection at XENON1T is given by $ \sigma_{a\to \gamma}^{A'\text{invPrim}} = 2  \sigma_{\gamma\to a}^{A'\text{Prim}}$. For $m_{A'} = 0.1~ (1) \,  \rm{keV}$,
\begin{equation}
\sigma^{A' \rm invPrim}_{a\to\gamma} = \sigma^{\rm invPrim}_{a\to\gamma}\times 400 (90) \times \frac{\alpha_B}{\alpha} \frac{g_{a\gamma A'}^2}{g_{a\gamma}^2}.
\end{equation}
Combining the solar axion flux and the detection cross section, we find that for $m_{A'} = 0.1 (1)$ keV, it requires \begin{equation}
g_{a\gamma A' } g_B \approx 0.11~(0.23) g_{a\gamma} e
\label{eq:sweet}
\end{equation} 
to explain the XENON1T excess.
Moreover, this choice of parameter helps to alleviate the HB cooling tension, such that its energy loss is reduced to 19\% (40\%) of that solely due to axion from Eq.~\ref{eq:axion}.

However, there are severe constraints for $U(1)_B$ couplings from astrophysics and collider physics.
The stars (the Sun, HB and supernova) can be cooled by directly emitting $A'$ through bremsstrahlung 
and Compton scattering. 
The constraint from SN1987 A for $U(1)_B$ is $g_B \lesssim 2.5\times 10^{-10}$ \cite{Rrapaj:2015wgs}. For solar and HB cooling, the emission of $A'$ from the ion leg is suppressed by $\mathcal{O}(m_e^2/m_n^2)$, and thus requires $g_B \lesssim 10^{-10}$ \cite{Harnik:2012ni, Hardy:2016kme}. For collider physics,  the UV anomaly cancellation of $U(1)_B$ leads to Wess-Zumino operators at low energy, which constrains $g_B/m_{A'} < 3\times 10^{-10} ~\text{keV}^{-1}$ \cite{Dror:2017nsg} from the invisible decays of $Z$-boson or mesons.
Therefore, both sets of constraints suggest that $g_B \lesssim 10^{-10}$ for keV $A'$.

To explain the XENON1T excess,
the coupling $g_{a\gamma A' } $ should be larger than $\sim 0.1 ~\text{GeV}^{-1}$, meaning a cutoff scale of $10$ GeV. Such energy scale may arise from integrating out new light particles~\cite{Paz:2020pbc}. 
However, the thermal photon in the plasma can decay via $\gamma^* \to a + A'$, thus a large $g_{a\gamma A' } $ is not desirable.
In summary, both $g_B$ and $g_{a\gamma A' } $ are highly constrained, such that they can not account for the excess.
One might also consider large $g_B$ to evade the cooling bounds, because the emitted $A'$ is then trapped inside the star.  Using a simple mean-free-path criterion, $(n_I \sigma_{A' I \to \gamma I})^{-1} \lesssim r_\text{star}$, where $\sigma_{A' I \to \gamma I}$ stands for the Compton scattering, the trapping is realized with $g_B \gtrsim 10^{-2}$ for the Sun and HB. 
$A'$ with a large $g_B$ can have a sizable coupling to electrons via 1-loop induced kinetic mixing, which is constrained by the $(g-2)_e$ measurement to be $g_e \lesssim 10^{-5}$~\cite{Knapen:2017xzo}. The loop induced coupling to electron is on the order of $g_e \sim g_B \alpha$. However, there can also be a UV contribution to the kinetic mixing as well. With fine-tuning, the total coupling to the electron can be made to be consistent with the constraint. The resulting scenario could have interesting implications which deserve further study.

We also consider using the environment effect to weaken the HB cooling bound while keeping the solar axion flux intact. The core temperature of HB and the Sun are $10$ and $1$ keV, and the corresponding photon plasma mass $\omega_p$ are $2$ keV and $0.3$ keV, respectively. Consider axions coupling to both photon and $U(1)'$ dark photon,
\begin{align}
{\mathcal L} \supset - \frac{1}{4} g_{a\gamma} a F_{\mu\nu} \tilde{F}^{\mu\nu}- \frac{1}{2} g_{a\gamma A'} a F'_{\mu\nu} \tilde{F}^{\mu\nu}  \ .
\end{align}
The resulting Primakoff production of axion contains diagrams with $t$-channel $\gamma$ or $A'$. One can choose the sign of 
$g_{a\gamma A'}$ to have destructive interference among the two diagrams and  make them cancel for HB with a particular combination of $ g_{a\gamma A'} g_B$ . However, a complete cancellation only works for particular momentum transfer 
and thus does not apply to the entire phase space. In addition, the core temperature for HB is not a constant, thus the cancellation may only happen within a restricted volume. Therefore, it is difficult to have a significant  environment-dependent suppression.

Above we focused on the solar axions, where the origin of the keV scale of the excess can be naturally explained. Nevertheless,
it is hard to accommodate the stellar cooling bounds. Dropping the relation between the keV scale and the solar axion energy, one may consider a boarder range of the signal sources. 
For example,  the $U(1)_B$ dark photon could be the dark matter, with mass $m_{A'} \simeq 2.8$ keV.
In this case, the detection at XENON1T is through the dark photon conversion $A' + \text{Xe} \to \gamma + \text{Xe}$ with the cross-section given by
\begin{align}
\sigma v= A^2 Z_{sc}^2 \frac{e^2 g_B^2}{6 \pi m_\text{Xe}^2},
\label{eq:dpdm}
\end{align}
where $A =131$, $Z_{sc} \sim 5.3 $. Since $A'$  is much lighter than nucleus, the photon energy is about $m_{A'}$. 
The total number of events is given by 
$
\frac{\text{exposure} }{m_\text{Xe}} \sigma v \frac{\rho_{DM}}{m_{A'}},
$
where $\rho_\text{DM} = 0.3 \,\text{GeV}/\text{cm}^3$ is the local DM density. The excess can be explained with $g_B \approx 5\times 10^{-11}$. Such value is marginally below that from the astrophysical and collider bounds $g_B \lesssim 10^{-10}$. Ref.~\cite{Aprile:2020tmw, 1802729} studied kinetic mixing dark photon dark matter as a solution for the excess and requires the mixing parameter $\epsilon \approx 7\times 10^{-16}$. The difference between those models and ours comes from the dark photon coupling to electrons. The resulting dark-photoelectric cross section is approximately $\frac{m_N^2}{m_e^2}\frac{\epsilon^2}{g_B^2}$ times of ~\eqref{dpdm}. Though there is a large difference in the value of couplings, both scenarios are marginally allowed given the astrophysical constraints.
\\

\noindent \textit{\textbf{Conclusions.}}---
Solar axion is an appealing explanation for the XENON1T excess, with its energy naturally in the keV range. 
In this letter, we have emphasized the importance of including photon with a similar recoil spectrum as a possible explanation for the XENON1T excess. In particular, it can significantly reduce the tension between the solar axion explanation and the astrophysical data, especially the stellar cooling bound. Introducing additional new physics can further alleviate the remaining tension. 

We conclude here by briefly discussing future prospects. We expect further sharpening the stellar cooling bound certainly helps to clarify the situation. If there is indeed additional new physics that helps to relieve the tension with the astrophysical bound, it would be interesting in exploring other possible signals of these new physics. For example, a more sensitive search for the  $U(1)_B$ can have the potential of shedding new light on this scenario. We also note that it is possible to have new physics models in which the photon comes from completely different sources. For example, it can come from a different dark matter scattering process \cite{Paz:2020pbc} or from decaying from an excited state of the dark matter \cite{Bell:2020bes, 1802727}, or the dark photon conversion process considered in this paper. In these cases, the spectrum of the photon would be different from the one from the inverse Primakoff process.  Future data can be used to distinguishing these scenarios. 
\\

\textbf{Note added:} Shortly after this work appeared on arXiv, a study \cite{Dent:2020jhf} appeared and also investigated the inverse Primakoff effect on the solar axion detection. 
Recently, Ref.~\cite{Abe:2020vff} reported that in the cross-section calculation, using the form factor from the relativistic Hartree-Fock (RHF) method led to a significant deviation from that based on a screened Coulomb potential. In previous version, we used the same RHF form factor data [54] to extract the screening length $r_0$. In this revision, we have reexamined our analysis and revised the extracted screening length. With this new value, the result from the screened Coulomb method agrees with that from the RHF method~\cite{Abe:2020vff}. The figures are modified accordingly and the coupling $g_{a\gamma}$ is shifted by a factor of $1.5 \sim 2$ in Fig.~\ref{fig:fit}. 

\begin{acknowledgments}
~\\
\noindent\textit{\textbf{Acknowledgments}}
We would like thank Luca Grandi, Tarek Saab, Evan Shockley for discussing in detail the response to the electron and photon in the XENON1T detector, Fei Gao, Jingqiang Ye for the details of fit and the analysis, Aaron Pierce and Sergey Sibiryakov for the axion models and 
constratins, and Maurizio Giannotti, Andreas Ringwald, and Oscar Straniero for clarifying the weighted average for $R$-parameter and the theoretical prediction formula. 
CG is supported by by Fermi Research Alliance, LLC under Contract No. DE-AC02-07CH11359,
JL acknowledges support by an Oehme Fellowship, 
LTW is supported by the DOE grant DE-SC0013642, 
XPW is supported by the DOE grant DE-AC02-06CH11357,
WX is supported by the DOE grant DE-SC0010296, and
YZ is supported by the Kavli Institute for Cosmological Physics at the University of Chicago through an endowment from the Kavli Foundation and its founder Fred Kavli.
\\
\end{acknowledgments}

\noindent \textit{\textbf{Appendix: Dependence of $R$-parameter constraints on the He abundance.}}---
The quickened core-burning process of HB and RGB from axion cooling can be compensated by a larger He abundance. This leads to a degeneracy between the He mass fraction, $Y_\text{He}$, and the axion couplings, $g_{ae}$ and $g_{a\gamma}$ when setting up constraints with observed $R$ parameter and weakens the coupling constraints when the uncertainties on the He abundance is large. 

The determination of $Y_\text{He}$ is particularly changeling for GC due to the absence of the spectroscopic window in the direct detection and the difficulties in stellar simulation. Given the similar O/H composition between the selected GCs and the low-metallicity HII regions, Ref.~\cite{Ayala:2014pea} uses the $Y_\text{He}$ of the later environment to approximate that of the former one and adopted $Y_\text{He} = 0.254\pm 0.003$. Ref.~\cite{Ayala:2014pea} also adopts the He abundance from the Big-Bang nucleosynthesis and that from the early solar system as the lower and higher bounds for $Y_\text{He}$ in GCs.   

\begin{figure}[htbp]
	\centering
	\includegraphics[width=0.48\textwidth]{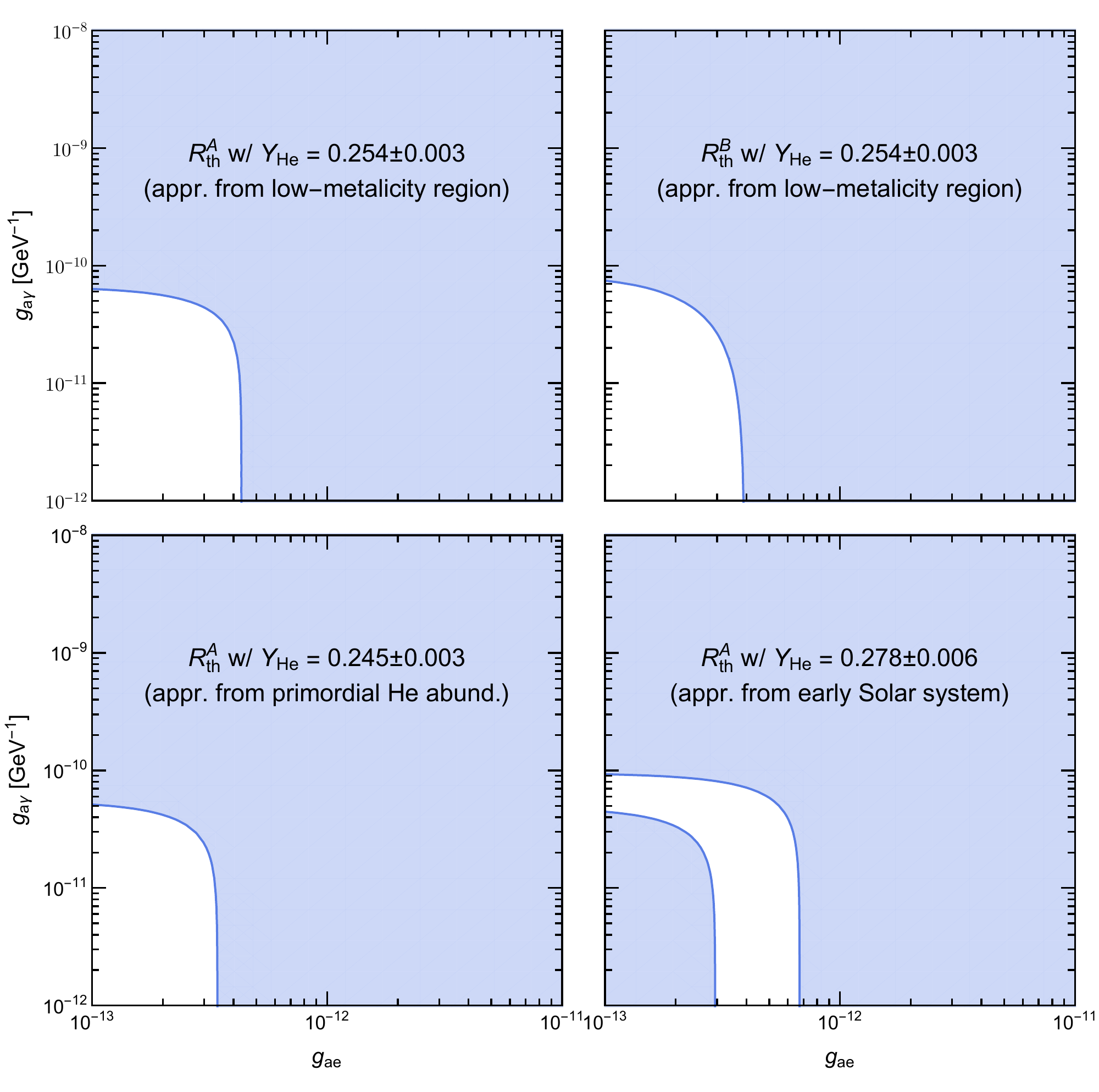} 
	\caption{95\% C.L. excluded parameter space (shaded with blue) from the ratio of the number of HB stars to that of RGB stars in GCs (a.k.a., $R$-parameter). Here we adopted the averaged $R$-parameter of $R_\text{av} = 1.39 \pm 0.03$~\cite{Ayala:2014pea} and considered the theoretical models of \eqref{rth1} and \eqref{rth2}. We approximate the $Y_\text{He}$ value from the low-metallicity region (upper),  the primordial He abundance (lower left) and the early Solar system (lower right).}
	\label{fig:Rpara}
\end{figure}

Ref.~\cite{Giannotti:2015kwo} updates the theoretical predictions of the $R$-parameter by including both the $g_{ae}$ and $g_{a\gamma}$ coupling from two stellar evolution models. The two predictions (labeled as $A$ and $B$) are given by 
\begin{align}
	R_\text{th}^{A} ={}& 6.26 Y_\text{He} - 0.41 \left(\frac{g_{a\gamma}}{10^{-10} \gev^{-1}}\right)^2 -0.12 \nonumber\\
	&- 0.0053 \left(\frac{g_{ae}}{10^{-13}}\right)^2 - 1.61 \delta \mathcal{M}_c \ , 
	\label{eq:rth1}
\end{align} 
or
\begin{align}
	R_\text{th}^{B} ={}& 7.33 Y_\text{He} - 0.095 \sqrt{ 21.86+21.08\left(\frac{g_{a\gamma}}{10^{-10} \gev^{-1}}\right)} \nonumber\\
	&+0.02 - 0.0053\left(\frac{g_{ae}}{10^{-13}}\right)^2 - 1.61 \delta \mathcal{M}_c \ , 
	\label{eq:rth2}
\end{align}
where
\begin{align}
	\delta \mathcal{M}_c\!=\!0.024\!\! \left[\!\left(\!\left(\!\frac{g_{ae}}{10^{-13}}\!\right)^2\!\!\!+1.23^2\!\right)^{\!\frac{1}{2}}\!\!-\!\!1.23\!-\!0.138 \left(\frac{g_{ae}}{10^{-13}}\right)^{\!\frac{3}{2}}\!\right].
\end{align}
In Fig.~\ref{fig:Rpara}, we showed the resulting~95 \% C.L. constraints on the $g_{ae}-g_{a\gamma}$ plane with the suggested value $Y_\text{He}=0.255\pm 0.03$ from the low-metallicity region~\cite{Ayala:2014pea}. To highlight the consequence of $Y_\text{He}$ uncertainty, we also set $Y_\text{He}$ of GCs to that of the primordial He abundance $Y_\text{He}=0.245\pm0.003$~\cite{Tanabashi:2018oca} and that of the early Solar system~\cite{Serenelli_2010}, $Y_\text{He}=0.278\pm 0.006$. Note that by approximating $Y_\text{He}$ to the early Solar system value, we assume no chemical evolution occurred during the 8 Gyr between the formation of GC and the Solar system. This is very unlikely. 

\bibliographystyle{utphys}
\bibliography{ref}

\end{document}